\documentstyle[12pt,sprocl,epsf]{article}

\def\be{\begin{equation}}
\def\ee{\end{equation}}
\def\beq{\begin{eqnarray}}
\def\eeq{\end{eqnarray}}
\def\T{{\cal T}}

\begin{document}
\title{DIFFUSING RECONSTITUTING DIMERS: A SIMPLE MODEL OF BROKEN
   ERGODICITY AND AGEING}

\author{DEEPAK DHAR} 

\address{Theoretical Physics Group, 
  Tata Institute of Fundamental Research, Homi Bhabha Road, Mumbai 
400005 (India)}     

\maketitle\abstracts{We consider a model of assisted diffusion of
hard-core particles on a line. In this model, a particle can jump left
or right by two steps to an unoccupied site, but only if the the site
in between is occupied.  We show that this model is strongly
non-ergodic, and the phase space breaks up into mutually disconnected
sectors. The number of such sectors increases as the exponential of the size 
of the system. Within a sector, the model can be shown to be equivalent to
the Heisenberg model, and is exactly soluble. We study how
time-dependent autocorrelation functions in the model depend on the
sector. We also show how this strong breaking of ergodicity ( which
can be thought of as an infinity of conservation laws) affects the
hydrodynamical description of the long wavelength density fluctuations
in the system. Finally, we study the effect of allowing transitions
between sectors at a very slow rate.}

\section{Slow and fast processes}

{\it If a system is weakly coupled to a heat bath at a given 
``temperature'', if the coupling is indefinite and not known precisely, if 
the coupling has been on for  a long time, and if all the ``fast'' things 
have happened and all the ``slow'' things not, the system is said to be in 
thermal equilibrium.
\hfill           R. P. Feynman \cite{feynman} 
}
\break

   The Boltzmann-Gibbs method of calculating properties of a 
system of large number of interacting particles by defining
a suitable ensemble of systems, and replacing the 
calculation of time-averages of obsevables by their phase 
space averages, is very powerful technique in statistical 
mechanics. To calculate the equilibrium properties of a 
system of a large number of degrees of freedom, all we have 
to do is to calculate the corresponding partition function,
and thus the free energy. All quantities of interest can 
then be determined by taking derivatives of this free energy 
with respect to suitable conjugate fields. Of course, this calculation is 
quite nontrivial, involves evaluation of many-dimensional integrals, and 
can be exactly done only for a very small class of soluble models. 
However, standard approximation techniques, such as weak- or strong- 
coupling perturbation expansions, or variational schemes can be used to 
get answers which are as close to the correct answer as desired away from 
critical points, and near critical points other schemes such as 
renormalization group, $\epsilon-$ expansion etc. can be used to 
determine the behavior of different quantities. Thus, it is perhaps not 
an exaggeration to say that the general principles of determining 
equilibrium properties of systems are quite well understood by now.

   However, many systems of interest, such as diamonds or 
ordinary glass, are not in thermal 
equilibrium, even though their properties do not seem to 
change significantly with time for the time-scale of 
experiments. These systems are said to be in metastable states.
It is well realized that these states have much higher free 
energy than some other stable state of the system, and if 
the system could easily explore all parts of the phase space 
available to it, such states would occur with negligible probability.
They owe their existence to the fact that the rate of 
transition to the stable phase can very low, and the system 
is {\it not ergodic}. 

   A precise microscopic specification of all positions and 
momenta of all the atoms, to describe the state of a small 
piece of glass is impossible. Clearly, we can, at best, hope to give a 
probabilistic description of the state. However, the ensemble 
of microscopic states having finite probability must to 
restricted to those which are `nearby', and are accessible to the system 
within times comparable to the experimental time-scales. Such an 
ensemble in these systems is much smaller than the corresponding 
microcanonical 
ensemble. It is easy to see that phase space volume of accesible states 
of such a system occupies a negligible fraction of the full 
phase space, and this fraction decreases as $exp(-a V)$, as the 
volume of system $V$ increases. The coefficient $a$ seems to have 
only a very weak dependence on the time scale used to define 
the set of 'accessible' states. Of course, $a$ must become 
zero as this time tends to the Poincare recurrence time of the 
system.

   However, it has proved difficult to define this concept
of restricted ensembles in a way which allows calculation
of such partition functions, and thus determine the average
properties of, even model systems. This is what we shall try 
to do here for a very simple toy model called the diffusing
reconstituting dimers (DRD) model. The ideas developed 
in this paper have been discussed in a recent paper \cite{gimenon}.
In this present paper, our aim is to argue that this is a very nice and 
tractable, hence useful model for 
glassy dynamics, while the earlier paper dealt more with its 
integrability, and connection to other soluble models.  We shall not 
attempt to review here the very large body of work which already exists 
dealing with the phenomena of ageing and relaxation in spin glasses 
\cite{others}. What is presented is a somewhat personal perspective.

   The key observation here is note that the  processes 
operating in a glass may be divided into two classes: 
fast and slow. Fast processes are those whose time scales 
are much shorter than that of observation, e.g. the 
vibratory motion of atoms about their mean positions. Slow 
processes are those involving rearrangement of atomic 
configurations, particle- or vacancy diffusion, creep etc.
To define a quasi-equilibrium state of glassy systems, we 
have to imagine that ``all the fast things have 
happened'', and the slow things have not even started. To 
be precise, we assume that the system dynamics is such that 
the slow processes {\it do not occur at all.} Then the 
dyanmics is {\it nonergodic}. 
\begin{figure}[htb]
\centerline{
        \epsfxsize=8.0cm
        \epsfysize=6.0cm
        \epsfbox{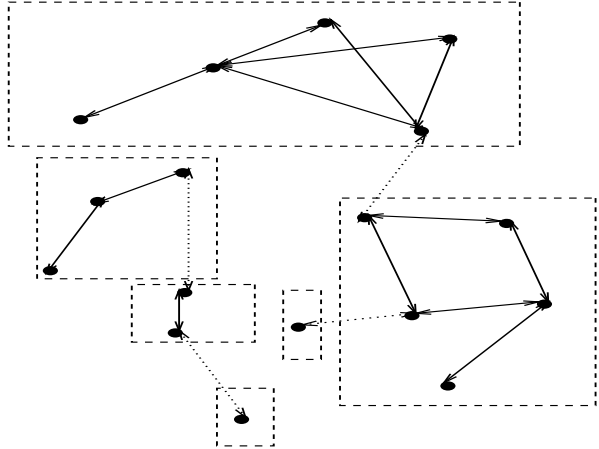}
}
\caption{A schematic representation of the decomposition of
phase space into disconnected sectors. The points denote different
configurations. Full lines show allowed fast transitions between
sectors. Different sectors are enclosed in rectangular boxes. Dotted
lines show slow transitions between different sectors.}
\label{fig:0}
\end{figure}
The phase space breaks up into 
a large number of disconnected pieces, called sectors. The 
number of such sectors increases exponentially with the size 
of the system. We shall call such system many-sector 
decomposable (MSD), and the ensembles which sum over only 
one of these sectors as {\it pico-canonical} ensembles.

   For the toy model we consider,  partition functions corresponding to 
such pico-canonical ensembles
can be explicitly computed. These differ from sector to sector. As the
number of sectors is of the order of $ exp (V)$, a precise specification
of the sector is too microscopic a detail, and not relevant. In an
experiment, one can, at best, hope to give some general probabilistic
characterization of the sector. This presumably would depend on the
history, and preparation of the system, e.g. cooling schedules etc.. In a
theoretical calculation, this implies that once the free energy in the
pico-canonical ensemble is determined, it must then be further averaged
over sectors with a suitable weight for each sector. 

   Finally, we shall show how we can describe the phenomena such as ageing
in glasses by taking into account the slow processes, as providing a
stochastic dynamics in which the system jumps from one sector to another
in time, and thus brings back ergodicity ( or tries to ) at very large
times. 
 
\medskip

\section{The Model}

     The model we shall choose to address these issues is a very simple
one. We consider $N$ hard-core point particles on a linear chain of $L$ 
sites.
There are no interactions amongst the particles except the hard-core 
interaction. We assume that the system evolves in continuous time time by a 
local Markovian dynamics. We shall choose the transition rates to satisfy 
the detailed balance condition, so that the rate from a transition, say from 
configuration $C$ to $C'$, is the same as the rate of the reverse transition
$C'$ to $C$. Then, in the steady state all accessible configurations occur 
with equal probability.

     Now for a more detailed specification of the allowed transitions. We 
assume that the particles can diffuse to nearby sites, but only if there 
is another particle nearby: a case of assisted diffusion. More precisely,
a particle at site i can jump left or right by two steps to the sites 
$ i{\pm}2 $, if that site is unoccupied, and the in-between site 
$i{\pm}1$ is occupied. If allowed, this transition occurs at rate 1.
    We represent the configuration of the system by a binary string 
$01100100$.... of length $L$, where the $i^{th}$ bit gives the occupation 
number 
$n_i$ of the site $i$. Then this process can be represented by a `chemical' 
equation
         $$ 110  {\leftarrow\rightarrow} 011$$

    An alternate way of viewing the system is to say that isolated 
particles cannot diffuse. But two particles at adjacent sites can 
diffuse together one step to the left or right with rate $1$. These pairs of 
particles will be called dimers. However, these pairings are not for ever, 
and the dimers can `reconstitute'. Thus, for example, in the sequence
of transitions   
$$    ..11010.. {\rightarrow} ..01110.. {\rightarrow} ..01011..$$
the middle particle is first paired with particle on the left, and then in 
the second transition with the particle on the right. We are thus 
considering a system of Diffusing Reconstituting Dimers (DRD).

    A third possible way of looking at the model is to think of the 
diffusion as that of 0's by two steps left or right if the intervening 
site is not a zero. This is then not assisted diffusion, but too many 
zeroes will `hinder' each other's diffusion. 

    The assumption in the model monomers have a much lower diffusivity than 
dimers is not easily realized in real physical systems. However, 
instances where this happens are not unknown: for example, platinum 
dimers on some surfaces have higher mobility than monomers.

\section{Many Sector Decomposition and Conservation Law of the 
Irreducible String}

 For of linear chain of L sites, the phase space consists of $2^L$ distinct 
configurations. It is however clear that the dynamical rules are such 
that the system is not ergodic. For example, the as the diffusion 
process conserves the number of particles, a system starting with N 
particles at some time will never be found in a configuration with a 
different number of particles. The phase space thus breaks up into 
mutually disconnected subsets, called {\it sectors}. The stochastic 
evolution of the system takes it from one configuration to another, within 
the same sector, but no transitions are allowed between different sectors.

 The conservation of particle number implies that the total phase space 
can be broken into $(L+1)$ disconnected subsets, each subset specified by 
a different value of the total particle number. Such conservation laws 
are well known in equilibrium statistical mechanics, and there effect is 
easily taken into account by defining suitably restricted ensemble : say
the canonical ensemble in which the number of particles is fixed. In 
addition, usually, one can prove that there is an equivalence of 
different ensembles ( except at special points like when the system is at 
a phase transition point). So one can even work with ensemble having a 
variable number of particles, so long as the chemical potential is chosen 
so that the average density comes out right.

 But this is not all. It is easy to see that if we break the linear chain 
into two sublattices (even and odd), then number of particles on each 
sublattice is separately conserved. Thus the phase space has to be broken 
up further. Most of the $(L+1)$ subspaces have to further subdivided into 
roughly L parts, so that the total number of disconnected regions of 
phase space is now of order $L^2$.  This would imply that to describe the 
long-time steady state of the system correctly, one would need not one, 
but two chemical potentials, one for each sublattice, and adjust them so 
that the initial state density on each sublattice is correctly 
reproduced.  
  
 So long as the number of conserved quantities is finite, one can 
continue this process, and eventually end up with a system described by
conventional Gibbs measure on a suitably defined `microcanonical' 
ensemble in the phase-space, or equivalently a `canonical' ensemble where 
the conservation laws are ignored, but their effects taken into account by 
including a finite number of thermodynamic potentials \cite{footnote1}. 
If there are m conserved quantities, and each 
takes order $L$ distinct values, the phase space has order $L^m$ distinct 
sectors.

 However, it is easy to see that in the DRD model, the breaking of  
ergodicity is much more severe. In fact the number of disconnected 
sectors increases as exponential of L. This is easy to see. Consider a 
configuration in which there are no adjacent sites both occupied such as
$$       ...0010100010010000101.....$$

   From the rules of the DRD dynamics, such a configuration cannot evolve at 
all, and remains the same at all times. The number of such configurations 
is easily shown to grow exponentially with $L$. [It is easy to
show that this grows as ${\mu}^L$,  where $\mu$ is the golden mean
=1.618...] 

\begin{figure}[htb]
\centerline{
        \epsfxsize=6.0cm
        \epsfysize=6.0cm
        \epsfbox{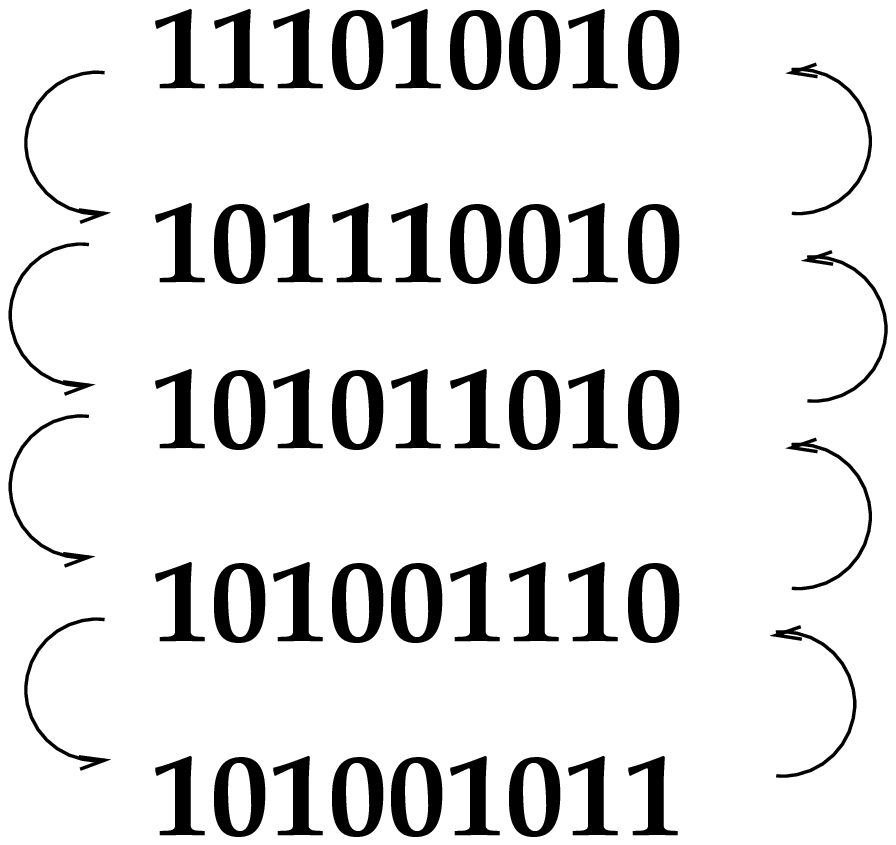}
}
\caption{ A non-trivial sector with L=9, and five states within the 
sector. The arrows indicate allowed transitions between these. For this 
sector the irreducible string is $1010010$.}
\label{fig:2}
\end{figure}


Consider now a less trivial case of single diffusing dimer in a frozen 
background.  This is shown in fig. 2. We see that while the movable $1$'s 
change in time, as the dimer reconstitutes as it moves, there is exactly 
one movable dimer in each such configuration if you want to move to the 
right,
and exactly one (possibly different) if you want a leftward move. After 
an order L moves in one direction, one reaches the end, and no further moves 
are possible. So there are about L configurations in each such sector.
The number of sectors again grows as ${\mu}^L$.

It is possible to continue this process, and construct sectors in which 
$2,3,..$ movable dimers are present. To get a full description of the 
phase-space decomposition, we need to develop a systematic method for 
doing this. This is most conveniently done in terms of a construction 
called the irreducible string (IS), which we describe next.

 With each of the $2^L$ possible configurations of the DRD model, we 
attach a binary string, called the IS corresponding to that 
configuration, constructed as follows: We take the $L$-bit binary string of
${n_i}$ specifying the configuration. We read this string from left to 
right until the first pair of adjacent $1$'s is encountered. This pair is 
deleted, reducing the length of the string by $2$. We repeat this process 
until no further deletions are possible. The resulting string is the IS
for the configuration. For example, for the binary string $01001110110$, the
irreducible string is $0100100$.

By construction, for each configuration, there is a unique IS. However, 
many different configurations may yield the same IS. The usefulness of 
this construction stems from the following observation: Two different 
configurations belong to the same sector, if and only if they both have 
the same IS. This is easy to see. 

 Firstly, we note that we need not assume that the pair of $1$'s to be 
deleted has to be the first occurrence of consecutive $1$'s when read 
from left to right. We get the same IS, whatever the choice and order 
of pairs $11$ deleted, so long as the final string has no pairs of $1$'s in 
it. [For example, from the configuration $101110$, we get the same IS $1010$,
whether the second and third $1$'s, or the third and fourth $1$'s are 
deleted.] Then, if we get configuration $C'$ from $C$ in one elementary step,
deleting the diffusing dimer first, we see that both $C$ and $C'$ have the 
same IS. Thus, the IS is a constant of motion.

 Secondly, if two configurations $C$ and $C'$ have the same IS, then there
exists a sequence of allowed transitions of the DRD model, which takes $C$
to $C'$.To prove this, notice that we can push any diffusing dimer in $C$ as
far right as possible, and get a {\it standard configuration} in the
sector, in which the configuration is the IS followed by all the dimers.
Thus, for example, the standard configuration corresponding to the 
configuration $101110$ is $101011$. Corresponding to the 
configuration $01001110110$, it is $01001001111$.
If $C$ and $C'$ have the same IS, they can both be transformed into the same
standard configuration, and as the DRD rules are reversible, into each 
other. So they belong to the same sector.

 The conservation of the IS thus gives us a convenient way to label the 
different sectors of the DRD model. To each allowed IS 
corresponds a unique sector. In addition, this decomposition of phase 
space into disjoint sectors using the IS as a constant of motion is 
maximal in the sense that {\it there are no other independent constants 
of motion}.

 As a side remark, let us note that the {\it number} of independent 
constants of motion is not a well-defined concept for a discrete system. 
For a 
classical mechanical system with n-dimensional phase space, we say that 
there are r independent constants of motion if the trajectories are 
confined to (n-r) dimensional manifolds. But for systems with discrete 
phase space, the trajectory consists of a finite number of 
discrete points, and its dimension is undefined.

  In fact, for such a system, given two 
independent constants of motion $I_1$ and $I_2$, one can always construct a 
third single constant of motion $I_3$, such that both $I_1$, and $I_2$ are 
known functions of $I_3$.  This result was proved in the context of quantum 
mechanics, by von Neumann  \cite{von} (for a recent discussion see Wiegert
\cite{wiegert}). His argument is quite simple, but  does not seem to be 
as well known as it should be.  Consider $I_1$, $I_2$ as two given commuting
quantum mechanical observables over
a finite dimensional Hilbert space. Since $I_1$ and $I_2$ are 
mutually commuting, they can be simultaneously diagonalized. Then the 
quantum-mechanical Hilbert space of all possible states of the system can be
represented as a direct sum of subspaces, each of which corresponds to a 
pair $(\lambda_1,\lambda_2)$, where $\lambda_1$ and $\lambda_2$ are 
eigenvalues of the operators $I_1$ and $I_2$. We order these subspaces in 
an arbitrary order, and define the operator $I_3$ to be block diagonal in 
this basis, with eigenvalue $r$ for all eigenvectors in the $r^{th}$ subspace.
Then, by construction, This operator commutes with both $I_1$ and $I_2$, and
its eigenvlaue $r$ tells us the sector, and hence both $\lambda_1$ and 
$\lambda_2$. Thus, the operators $I_1$ and $I_2$ are functions of $I_3$.

        Using the von Neumann construction, we can take any set of 
commuting constants of motion $I_1,I_2,I_3,...$, and collapse them into a 
single constant. In general, if one can construct an observable which takes 
the same value for all configurations within a sector, but distict values 
in distinct sectors, then such an observable is a constant of motion 
and there can not exist any other constants of motion independent of it.  
Conversely, a single constant of motion having $2^n$ 
distinct eigenvalues can be thought of as n {\it independent} constants 
of motion, each having only two distinct eigenvalues.

        For the DRD model, this implies that the IS can be thought of as 
a single constant of motion, equivalent to a large number of independent
constants of motion: If we count the number of 1's between the $i^{th}$ 
and $(i+1)^{th}$  zero in the binary string corresponding to the 
configuration, with time this number can only change in multiples of 2, 
hence its parity is conserved. If one is uncomfortable with constants of 
motion whose values are character-strings, the IS can be thought of as a 
binary integer.   

        There is another way to represent the IS as a constant of motion, 
which makes contact with the theory of integrable models. We consider two 
noncommuting matrices  $A(0)$ and $A(1)$ ( these can in general be 
complex  $n \times n$ martices, 
but may be taken to be $2 \times 2$ real matrices for our purpose here) which 
satisfy the relation
\begin{equation}
A(1) A(1) A(0)= A(0) A(1) A(1) 
\end{equation}
For any configuration $C$ of the DRD model with occupation numbers 
${{n_i}}$, we define a matrix $I_L(C)$ by
\begin{equation}
I_L(C) = \prod_{j=1}^L A(n_j) 
\end{equation}
where the matrix product is ordered so that larger $j$ values are more to 
the right. Then, from the eq(1) it follows that this matrix product does 
not change by an elementary evolution step of the DRD model. This implies 
that $I_L$  will not change in time even as $C$ changes, and thus is a 
contant of motion. Eq(1) does not completely determine $A(0)$ and $A(1)$, as
these forms only $4$ equations for $8$ real parameters making the
marices $A(0)$ 
and $A(1)$. Thus we can choose a one parameter family of matrices 
$A(0,\lambda )$ and $A(1,\lambda )$ depending on a real parameter $\lambda$.
A simple specific choice is
\begin{equation}
 A(0,\lambda)= \left[  \begin{array}{clcr}
              1- \lambda   &     \lambda  \\
                1           &    -1  
\end{array}  \right]
\end{equation}
and
\begin{equation}
   A(1,\lambda) = \left[ \begin{array}{clcr}
            1 & 0  \\
            0 & -1 
\end{array} \right] 
\end{equation}
Now $I_L$ becomes a 2X2 matrix of the form
\begin{equation}
    I_L(C,\lambda) =\left [\begin{array}{clcr}
                     I_{11}(\lambda) & I_{12}(\lambda)    \\
           I_{21}(\lambda)  & I_{22}(\lambda) 
\end{array} \right]
\end{equation}
whose elements $I_{11}, I_{12}, I_{21}$ and $I_{22}$ are 
polynomials in $\lambda$
of degree at most $L$. Since this matrix is conserved for all values of 
$\lambda$, it follows that all the coefficients of these polynomials
are constants of motion. These coefficients in terms can be written
out explicitly as functions of ${n_i}$.

\section{Equivalence to a Multi-Species Exclusion Process} 

   The algorithm to determine the IS also specifies the positions of the 
characters in the L-bit binary string that remain undeleted. As the 
configuration evolves, these positions will change, but the relative 
order of the characters remains unchanged. The DRD model can thus be 
viewed as a process of diffusion of hard-core particles on a line. Let 
the IS be of length $\ell$. Then the number of particles is $\ell$. Let
$X_i(t)$ be the position of the $i^{th}$ character of the IS (counting 
from left). Each particle carries with it a `spin label' which is 0 or 1,
and is unaffected by the dynamics. Thus, our model is a special case of
the Hard-Core Random Walkers with Conserved Spin (HCRWCS) models which 
may be defined for all 1-d models where a conserved IS can be defined
\cite{dhar,PRL,harimenon,thesis}.

  If the $i^{th}$ character of the IS is a 1, it must be followed by a 0 in 
the IS, and also on the chain, this character must be immediately after it.
so that we have
$$    X_{i+1}(t) = X_i(t) + 1 $$
for all times t, and the $i^{th}$ and $(i+1)^{th}$ walkers always move 
together. In this case it is better to think of these as a single walker,
which occupies two consecutive sites on the chain. Then there are two 
kinds of walkers: walkers with spin label 0 occupying  a single site, and 
walkers with label 10 which occupy two sites. In addition there are the 
sites not occupied by the random walkers, which are occupied by dimers 11,
which again occupy two sites. Let us call these particles of type
$B$,$C$ and $A$ respectively.
     We now show that the DRD dynamics can be viewed as a stochastic 
dynamics of a system of 3 species $A$, $B$ and $C$ of particles on a line. In 
the latter process ( to be called the Exclusion process (ExP)), each site 
of linear chain is occupied by  a single particle, which may be of type 
$A$, $B$, or $C$. We set up a one-to-one correspondence between the 
configurations of the DRD model, and of the ExP as follows: Read the 
binary string of the DRD from left to right. If the next character of the 
DRD model is 0, another character in the DRD string is read.We write a 
single character $A$ or $B$ for the ExP string, depending on if the last 
chacter read was $1$ or $0$. Conversely, the binary string 
corresponding to a configuration $ABBCAABCB$... of the ExP is obtained by a 
simple substitution $A \rightarrow 11$, $B \rightarrow 10$, $C \rightarrow 
0$.  

   We can now write the stochastic evolution rule in terms of the ExP 
configurations. These are quite simple: A particle of 
type $A$
can exchange position with a particle of type
$B$ or $C$ at an adjacent site  with rate 1. Type $B$ and $C$ 
cannot exchange positions with each other. 
This model thus is  a special case of the k-species exclusion process 
defined by Boldrighini et al \cite{boldri}. In the latter, there are k 
distinct species of particles (k=2,3,4..). and particles can exchange 
places with adjacent particles with  a rate which depends on both the 
species of the particles beinf interchanged. In our case k=3. The 
k-species model has been studied recently Derrida et al \cite{derrida},
and by Ferrari et al \cite{ferrari}.

   The conservation of IS in this language corresponds to the simple
statement that as $B$ and $C$ type particles cannot exchange places, their
relative order is unchanged in time. Consider a string composed of three
characters $A$, $B$ and $C$ which specifies a configuration of the ExP
process.  Then deleting all the occurrences of $A$'s in the string, we are
left only with a string composed of $B$'s and $C$'s which is conserved by
the dynamics.

\section{Calculation of the Number and
Sizes of Sectors}
    
    In terms of the ExP model, it is quite straightforward to write  down 
the formulas for the number of distinct sectors for the DRD model, and 
also the number of configurations in each sector.
     Consider free boundary conditions for convenience. In a sector having
$N_A$ particles of type $A$, $N_B$ particles of type B, and $N_C$ particles of 
type $C$, the total number of sites is $L=2N_A +2N_B +N_C$. The number of 1's
in this case is $2N_A + N_B$, and the number of 0's is $N_B + N_C$. The 
length of the IS  is $2N_B + N_C$.

 The number of 
distinct sectors is the number of distinct ways of writng the IS 
consisting of $N_B$ $B$'s and $N_C$ $C$'s. Since all choices are allowed, this is 
$ ( N_B + N_C)! /(N_B ! N_C! )$. The number of configuration within a sector
with a specified IS, say  $BCCCBBCBC$...., is the number of ways we can 
place $N_A$ $A$'s between $(N_B + N_C)$ $B$'s and $C$'s. So this number is
\begin{equation}
    \Omega (N_A,N_B,N_C) = (N_A + N_B+ N_C)/[ N_A! (N_B + N_C)!] 
\end{equation}
 This determines the picocanonical entropy in each sector.
   To calculate the total number of sectors, we have to sum the number 
of sectors over all 
allowed values of $N_B$ and $N_C$. 

  Multiplying $\Omega (N_A,N_B,N_C)$ by the number of sectors with $N_B$
B-type and $N_C$  C-type particles, we get the the total number of 
configurations having $N_A,N_B$ and $N_C$ particles of type A,B,C 
respectively. Denote this by $\Xi(N_A,N_B,N_C)$.
Then we get
\begin{equation}
\Xi (N_A,N_B,N_C) = (N_A + N_B + N_C)!/[N_A! N_B! N_C!]
\end{equation}
  Extremizing this with respect to $N_A, N_B$ and $N_C$, subject to the 
constraint that $2N_A + 2N_B + N_C=L$, we see that the this is extremized 
for $N_A= N_B= N_C/2 = L/6$.  Hence if we pick a configuration at random, 
its length is most likely to be approximately $2L/3$, and it will have 
approximately $L/6$ diffusing dimers. The typical number of 
configurations in this sector is ${\kappa}^L$ where ${\kappa} = 
2^{4/3}.3^{-1/2}$. The number such sectors is approximately ${\kappa}'^L$, 
where ${\kappa}' = 3^{1/2}.2^{-1/3}$. [ We check that ${\kappa\kappa}'=2$,
so that the total number of configurations increases as $2^L$, as it must.]

  Alternatively, we can proceed directly as follows: 
Let $F_{\ell}(0)$ and $F_{\ell}(1)$ be the number of distinct strings of 
length $\ell$ which end in a $0$ and a $1$ respectively. Since in the IS, a 
$1$ must be followed by $0$, by a $0$ may be followed by either $0$ or $1$, 
these F's satisfy the recursion relations
\begin{equation}
F_{\ell +1}(0)=F_{\ell}(0) +F_{\ell}(1) 
\end{equation}
\begin{equation}
F_{\ell+1}(1)=F_{\ell}(0).
\end{equation}
Given the starting values 
$F_1(0)=F_1(1)=1 $
these recursions are easily solved, and we see the $F_{\ell}(1)$ is the 
${\ell}^{th}$ Fibonacci number which grows as ${\mu}^{\ell}$, where $\mu = 
(\sqrt 5 +1)/2$. For  a line of length L, the largest possible value of 
$\ell$ is L. A sector with $r$ diffusing dimers corresponds to $\ell =L-2r$.
Hence the total number of sectors is
\begin{equation}
N_L = \sum_{r=0}^{[L/2]} [F_{L-2r}(0) + F_{L-2r}(1)]
\end{equation}
where $[L/2]$ is the largest integer not greater than $L/2$.
This shows that the total number of sectors also increase as ${\mu}^L$
for large $L$. 

  For the DRD model, the probability weights in the steady state 
correspond to the  rather trivial Hamiltonian H=0, and so the
calculation above does not seem to have much interesting structure, or
variation from sector to sector. But it does provide us with a simple
system where one can completely characterize different sectors, and
calculate phase-space averages within a specified sector. 

\section{The Rate matrix as a Quantum Spin Hamiltonian}

The technique of writing down a quantum-spin hamiltonian corresponding to 
a Markov process for  a discrete system is by now quite well-known 
\cite{alc}. For completeness, we recapitulate it here briefly.  Let $P({\cal 
C},t)$ be
the probability that a classical system undergoing Markovian evolution
is in the configuration ${\cal C}$ at time $t$.  These probabilities 
evolve in time
according to the master equation
\be
{\partial P ({\cal C},t) \over \partial t} = \sum_{{\cal C}'} 
\left[-W ({\cal C}
\rightarrow {\cal C}')\, P({\cal C},t) + W({\cal C}' 
\rightarrow {\cal C})\, P({\cal C}',t)\right]
\ee
where the summation over ${\cal C}'$ is over all possible 
configurations of the system
and $W({\cal C} \rightarrow {\cal C}')$ is the transition rate 
from configuration ${\cal C}$ to
configuration ${\cal C}'$.

We define a Hilbert space, spanned by basis vectors $|{\cal C}\rangle$,
which are in $1$ to $1$ correspondence with the configurations
${\cal C}$ of the system.  A configuration with probability weight 
$P({\cal C},t)$ is
represented in this space by a vector
\be
|P(t)\rangle = \sum_{{\cal C}} P ({\cal C},t) | {\cal C}\rangle
\ee
The master equation can then be written as
\be
{\partial |P(t)\rangle \over \partial t} = - {\hat H} |P(t)\rangle
\ee
This equation can be viewed as an imaginary-time Schodinger equation
for the evolution of the state vector $|P(t)\rangle$ under the action
of the quantum Hamiltonian ${\hat H}$.

The $2^L$ configurations of the DRD model on a line of length $L$ can be 
put in one-to-one
correspondence with the configurations of a spin-$1/2$ chain of $L$
sites.  At the site $i$, the spin variable is $\sigma_i^z = 2n_i - 1$, 
and takes values $+1$ and $-1$ if $n_i$ is $1$ or $0$ respectively.  It is
then straightforward to write ${\hat H}$ as the Hamiltonian of the spin
chain in terms of the Pauli spin matrices $\sigma_i$.  We find that
${\hat H}$ has local 2-spin and 3-spin couplings, and is given by
\be
{\hat H} = -\frac{1}{4}\sum_i \left[\vec\sigma_{i-1} 
\cdot \vec\sigma_{i+1} - 1\right] \, (1+\sigma^z_i)
\ee
This Hamiltonian is quite similar in structure to  an integrable
spin model with three-spin couplings proposed and solved by 
Bariev\cite{bariev}.
The Bariev model has the Hamiltonian given by
\be
\hat H_B = - {1\over 4} \sum_i(\sigma^x_{i-1} \sigma^x_{i+1} +
\sigma^y_{i-1} \sigma^y_{i+1}) (1 - U\sigma^z_i)
\ee
The sign of $U$ can be changed by the transformation $\sigma^x_i
\rightarrow \sigma^y_i$, $\sigma^y_i \rightarrow \sigma^x_i,$ and 
$\sigma^z_i \rightarrow
-\sigma^z_i$.  We set $|U| = 1$.  Then the Bariev model differs from
our model through the term
\be
\hat H - \hat H_B = H' = - {1\over 4} \sum_i 
\left(\sigma^z_{i-1}\sigma^z_{i+1} -
1\right) \left(1+\sigma^z_i\right)
\ee
It is easy to see that these terms commute with the IS 
operator $\hat I_L(\lambda)$.  Thus $\hat I_L(\lambda)$ also commutes
with $\hat H_B$, and provides an infinity of constants of motion of the
Bariev model (in the special case $|U| = 1$).

It is useful to write this Hamiltonian in terms of fermion operators
$c^+_i$ and $c_i$ defined by the Jordan-Wigner transformation
\beq
c^+_i & = & \exp \left[-i\pi \sum^{i-1}_{j=1} \sigma^+_j
\sigma^-_j\right] \, \sigma^+_i \nonumber \\
c_i & = & \exp \left[i\pi \sum^{i-1}_{j=1} \sigma^+_j
\sigma^-_{j}\right]\, \sigma^-_i
\eeq
In terms of these fermion operators, ${\hat H}$ can be written as
\be
{\hat H} = \sum_i \left[c^+_{i-1} n_i c_{i+1} + c^+_{i+1} n_i c_{i-1} + 
n_i (n_{i-1} - n_{i+1})^2\right]
\ee
where $n_i = c^+_ic_i$ is the number operator at site $i$.  The first
two terms of this Hamiltonian represent assisted hopping over an
occupied site, and the last term describes two- and three- body
potential interaction between nearby sites.  Similar hopping terms are
encountered in a model studied by Hirsch in the context of hole
superconductivity\cite{hirsch}.

\section{Conservation laws of the Quantum Hamiltonian}

While the evolution of probabilities in a
classical Markov process could be cast as the evolution of the
wavefunction of a suitably defined quantum mechanical problem,
the concept of constants of motion is quite different in these two cases,  
as I explain  below.

A classical observable ${\cal O}$ is a function which assigns a value, 
say in terms of a real
number ${\cal O} ({\cal C})$ to each possible configuration ${\cal C}$ of the 
system.  We say
that ${\cal O}$ is a constant of motion when the value of the observable
${\cal O}({\cal C}_t)$, associated with the configuration of the system
${\cal C}_t$ at any arbitrary time $t$,
is independent of $t$.  In a quantum mechanical formulation, observables
are represented by matrices $\hat{\cal O}$, and the quantity of interest
is the expectation value of the observable ${\cal O}$ at time $t$,
given by the formula
\be
\langle \hat{\cal O}\rangle_t = \langle1|\hat{\cal O}|P(t)\rangle,
\ee
where $\langle 1|$ is the row vector with all entries 1.  Clearly, in
this language, a classical observable ${\cal O}$ corresponds to a
quantum mechanical operator $\hat{\cal O}$, which is diagonal in the
natural basis of the system, i.e. when the basis vectors are
$|{\cal C}\rangle$.

We shall call a quantum mechanical observable $\hat{\cal O}$ a constant
of motion, if the matrix $\hat{\cal O}$ commutes with the Hamiltonian 
${\hat H}$.
In this case it is easy to verify that its expectation value evaluated 
with respect to the
probability vector $|P(t)\rangle$ does not change in time.  (Note that
we are using a different prescription for evaluating expectation
values than in standard quantum mechanics.)  Clearly the
class of all possible quantum mechanical constants of motion (all
matrices $\hat{\cal O}$) is much larger than that of the classical constants of
motion (diagonal in the configuration basis).  

      We have already seen that the
IS is a classical constant of motion.  Using its definition,
(Eq. 4.5), we define the diagonal matrix $\hat
I_L(\lambda)$ by
\be
{\hat I}_L(\lambda) = \prod^L_{j=1} A({\hat n}_i,\lambda),
\ee
where ${\hat n}_j = \sigma^+_j\sigma^-_j$ is now an operator, and not
a {\it c-}number.  It is easy to verify that
${\hat H}$ commutes with ${\hat I}_L(\lambda)$ , for all values of 
$\lambda$.
Thus ${\hat I}_L(\lambda)$ provides us with a one parameter family of
constants of motion.  In fact, as discussed earlier, there  are no 
additional independent classical constants of motion in our model. 

   As ${\hat H}$ commutes with ${\hat I}_L(\lambda)$, it has a
block-diagonal structure, with each block corresponding to a sector of
the phase space, and to a particular IS.  The
task of diagonalizing ${\hat H}$ then reduces to that of diagonalizing it in
each of the sectors separately.
 
   Consider any one of these sectors.  Let the IS ${\cal I}$ for
this sector, in the 3-species exclusion process notation, be a string of
$N_B\, B$'s and $N_C\, C$'s in some order, say ${\cal I} = BCBBCBCC \ldots$
The number of diffusing dimers in this sector then is $N_A= 
(L-2N_B-N_C)/2$.  A
typical configuration in this sector is specified by a string of
length $(N_A + N_B + N_C)$ with $N_A\, A$'s interspersed between the 
characters of the IS, e.g. BCABAABCBCAC $\ldots$.

We define a chain of spin-1/2 quantum spins ${\tau_i}$, with $i$ ranging
from $1\/$ to
$L_\tau = (N_A + N_B + N_C)$.  For each configuration in the sector ${\cal 
I}$, we
define a corresponding configuration of the $\tau$-spin chain by the
rule that $\tau_j^z = +1$ if the $j^{\rm th}$ character in the string
specifying the configuration is $A$.  For $\tau^z_j = -1$, 
the corresponding
character can be either $B$ or $C$.  But this degeneracy is completely
removed by using the known order of these elements in the irreducible
string ${\cal I}$.  Thus there is a one-to-one correspondence between the
configurations of the DRD model in the sector ${\cal I}$, and the
configurations of the $\tau$-chain with exactly $N_A$ spins up.

The action of the Hamiltonian $\hat H$ on the subspace of configurations
in the sector ${\cal I}$ looks much simpler in terms of the $\tau$-spin
variables. It is easily checked that the evolution 
in terms of the $\tau$-variables is that of a simple exclusion process. 
The quantum Hamiltonian for this process is the well known Heisenberg
Hamiltonian
\be
{\hat H}_{Heis} = - \sum^{L_\tau - 1}_{i=1} \left(\vec \tau_i \cdot \vec
\tau_{i+1} - 1 \right).
\ee
Note that ${\hat H}_{Heis}\/$ looks different for sectors
with differing $\L_\tau$'s. We have defined the transformation between 
the $\sigma-$ and $\tau-$ spin variables by an algorithmic prescription. 
It seems very difficult to do so by an explicit formula. The 
transformation is very complicated, and non-local.

We would now like to construct a one-parameter family of operators
that commute with ${\hat H}_{Heis}$.  This is a well-known construction for the
Heisenberg model \cite{panchgani}.  We use periodic boundary conditions for
convenience.  Define $2 \times 2$ matrices $L_j(\mu)$ whose elements
are operators acting on the spin $\tau_j$, and $\mu$ is a parameter
\be
L_j(\mu) = \Biggl( \matrix{1 + i\tau_j^z \mu & i\mu \tau^-_j \cr i\mu
\tau_j^+ & 1 - i\tau^z_j\mu} \Biggr),
\ee
and define
\be
\hat \T(\mu) = Tr \prod^{L_\tau}_{j=1} L_j(\mu).
\ee
Then it can be shown that \cite{panchgani} for all $\mu,\mu'$
\be
\left[\hat \T (\mu), \hat \T(\mu')\right] = 0.
\ee
Writing $\ln \hat\T(\mu) = \sum^\infty_{r=1} \mu^r \hat J_\mu$
we get $\left[\hat J_r, \hat J_s\right] = 0$, for all $r,s$ and $\hat
J_2 = {\hat H_{Heis}}$.  Thus the set of operators $\{\hat J\}$ constitute a
set of quantum mechanical constants of motion for the Hamiltonian
${\hat H}_{Heis}\/$ in each sector separately, and hence for ${\hat H}\/$.

The Hamiltonian ${\hat H}$ has still another additional infinite set of
constants of motion.  These are related to the existence of an
additional symmetry in the model.  Clearly, replacing a $B$-type
particle by a $C$-type particle does not affect the dynamics of the
exclusion process.  Thus, the full spectrum of eigenvalues of ${\hat H}$
in any two sectors of the DRD model with different irreducible
strings, but having the same values of $N_A$ and $N_B + N_C$, is exactly the
same. Such a symmetry may be viewed as a local gauge symmetry between
the $B$ and $C$ `colors'.  Changing the character in the
IS from $B$ to $C$ (or vice versa) changes the length
of $L$ by 1.  This symmetry therefore relates two sectors of the DRD 
model with different sizes of the system. 

The simplest inter-sector operators which preserve the
total length of the chain $L$ are operators which interchange two
characters of the IS.
Let ${\hat K}_j$ be the operator which interchanges the $j^{\rm th}$
and $(j+1)^{\rm th}$ characters of the IS.  Then
clearly, we have
\be
\left[{\hat H},{\hat K}_j\right] = 0 \,\, {\rm for}\,\, j = 1 \,\, {\rm
to} \,\, L_\tau - 1.
\ee
  This
`color' symmetry implies that the spectrum of ${\hat H}$ is the same in
all the sectors with the same value of $L$ and $N_A$.

We have thus shown that the DRD hamiltonian shows the existence of three 
different classes of infinity of constants of motion: the 
classical-mechanical constants of motion which may be encoded in the 
conservation of IS, the off-diagonal constants of motion of the 
Heisenberg model which can be used to diagonalize the martix ${\hat H}$ 
within a sector, and the constants of motion ${\hat K}$ which are also
off-diagonal, but relate eigenvalues and eigenvectors in different sectors.
 
\section{Time-dependent Correlation Functions}

The static correlation functions in the DRD model are quite 
trivial because of the simple form of the hamiltonian 
chosen. The effect of ergodicity- breaking and constrained 
evolution is best seen in the time- dependent density autocorrrelation 
functions. These show interesting 
variations from one sector to another. Such variations occur despite the
fact that, in each sector, there is a mapping between the DRD model and
the simple exclusion process whose dynamics is known to be governed
by diffusion. As we will see below, this mapping leads to a 
correspondence between tagged hole correlation functions in the
two  problems, but the form of autocorrelation function decays depends 
on the IS, and can be quite different.

Consider a particular sector with IS ${\cal I}$. Let the number
of diffusing A, B and C particles in the equivalent 3-species exclusion
process be $N_A, N_B$ and $N_C\/$ respectively. In the IS, let $b(k)$ be the 
number of B's to the left of the k-th hole.( A hole is a particle of type 
$B$ or $C$.) Evidently, the function $b(k)$ specifies the IS completely. 

Different configurations in the sector are obtained from different 
distributions of A's in the background of B's and C's. If there are
$a_k$ A's to the left of the k'th hole, the location of this
hole in the ExP  and DRD problems is
\begin{eqnarray}
y_{ExP}(k,t) &=& k + a_k(t), \\   
y_{DRD}(k,t) &=& k + b(k) + 2 a_k(t).
\end{eqnarray}
Notice that $b(k)$ (unlike $a_k$) is time-independent. Defining a
tagged-hole correlation function for the ExP and DRD problems in 
analogy with the conventional tagged-particle correlation 
function\cite{majbar} through
\begin{equation}
\sigma^2(t) = <[y(k,t) - y(k,0)]^2>,
\end{equation}
we see that the simple relationship
\begin{equation}
\sigma_{DRD}(t) =  2 \sigma_{ExP}(t)
\end{equation}
holds.  However, no simple, exact equivalence between the two problems 
can be established for tagged-{\it particle} correlations or single-site
autocorrelation functions. This is because the
transformation between spatial coordinates in the DRD and related 
ExP problems is nonlinear; a fixed site in the former problem corresponds to
a site whose position changes with time in the latter.

Let $x\/$ and $\xi\/$ denote spatial locations in the DRD and ExP
problems respectively and let $a_\xi$ be the number of A's to the 
left of $\xi$.  Evidently, the number of holes (B's and C's) to
the left of this site is $(\xi - a_\xi)\/$. In the equivalent
configuration in the DRD problem, each  A corresponds to two
particles, while the $(\xi -  a_\xi)\/$ B's and C's occupy a length
$b(\xi - a_\xi) + \xi - a_\xi\/$ which depends on the IS in question. 
Thus
\begin{equation}
x = a_\xi + b(\xi - a_\xi) + \xi,
\end{equation}
while the number of particles $a_x$ to the left of $x$ is given
by 
\begin{equation}
a_x =  2 a_\xi + b(\xi - a_\xi).
\end{equation}
The transformation between the integrated particle densities 
$a_x$ and $a_\xi$ is therefore quite complicated. It depends on the IS
through the function $b\/$ and is highly non-linear.

The correlation functions involving $a_\xi$ are quite simple. The
density-density correlation function for the ExP in steady state is 
defined as \begin{equation}
C_{ExP}(\xi,t) = <n^\prime(\xi_0+\xi,t_0+t)n^\prime(\xi_o,t_0)>.
\end{equation}
where $n^\prime$ is a particle occupation number and an average over
$\xi_0$ and $t_0$ is implicit. $C_{ExP}$ satisfies the simple
diffusion equation.
\begin{equation}
\frac{\partial C_{ExP}(\xi,t)}{\partial t} = \nabla^2_\xi C_{ExP}(\xi,t),
\end{equation}
where $\nabla^2_\xi$ is the discrete second-difference operator. For large $t$,
therefore, $C_{ExP}$ decays as $t^{-1/2}$. Since $a_\xi$ is a space integral
over $n^\prime(\xi)$, correlation functions involving $a_\xi$'s can
be obtained as well. However, the change of variables from $a_\xi$ to
$a_x$ (Eqs. 8.5 and 8.6) is difficult to perform explicitly.
Nevertheless, we can determine the asymptotic behaviour of correlation
functions $C_{DRD}(x,t)$, defined analogously to Eq. 8.7.
  Our technique for determining the long-time behavior of these correlations
has been used earlier in discussing other models of the HCRWCS type 
\cite{PRL,harimenon,chapter,intcol}. It relies on the 
fact that the slowest modes in the problem are diffusive motion of 
characters of IS, and the persistence of local density fluctuations of $1$'s
determines the decay of autocorrelation function at large times. We  
illustrate the technique below for various sectors. All these 
conclusions have been verified by extensive Monte Carlo 
simulations\cite{gimenon}.

The simplest sector is the one characterized
by the IS $101010 \ldots$ of length $\ell$
which is a finite fraction of the total length $L$.  In this sector, 
all the odd sites
are always occupied and the dynamics on the even sites is that of the
simple exclusion process. It is known that in the latter, the 
density-density autocorrelation function has a $\tau^{-1/2}$ tail for large 
times $\tau$. Thus $C_{DRD}(x_0,\tau)$ 
is zero for $x_0$ on the odd sublattice, while it decays as $\tau^{-1/2}$ on
the even sublattice.

Next consider the case where the IS consists of
all zeros $0000\ldots$, whose length is a nonzero 
fraction of
$L$. Again, the $t^{-1/2}$ decay of $C_{ExP}$  then implies a similar 
behaviour for $C_{DRD}$.

Now consider the general case of an IS whose
$j^{\rm th}$ character is $\alpha_j = 0,1$.  Let us assume that at time
$t_0$, the site $x_0$ is occupied by a particular character of the IS. 
Between the time $t_0$ and $t_0+t$,  let the net number of dimers which cross
the point $x_0$ towards the right be $m$ (a leftward crossing being counted 
as a contribution $-1$ to $m$).  For large times $t$, it
is known that the distribution of $m$ is approximated by a Gaussian
whose width increases as $t^{1/4}$, i.e.
\be 
{\rm Prob} (m|t) \approx {1 \over \sqrt{2\pi}\Delta_t} \exp
\left[{-m^2 \over 2\Delta^2_t}\right].
\ee
where $\Delta_t$ increases as $t^{1/4}$ for large $t$\cite{arratia}.  
If $m$ dimers
move to the right, a site occupied by $\alpha_j$ at time $t_0$ will now be
occupied by $\alpha_{j+2m}$ at time $t_0+t$.  Hence the autocorrelation function
$C(\tau)$ is approximated by
\be
C(\tau) = \sum^\infty_{m=-\infty} \overline{\alpha_j\alpha_{j+2m}}
\cdot {\rm Prob} (m|t),
\ee
where $\overline{\alpha_j\alpha_{j+2m}}$ is the average value of the
correlations of characters in the IS averaged over
$j$.  This can have different values for even and odd $j$'s as
a particular element of the IS always stays on one
sublattice.  Define
$\gamma_{\buildrel {\rm odd} \over {\rm even}}  (m) = 
\overline{\alpha_j\alpha_{j+2m}}$, averaged over odd/even sites.  
We have
\be
C_{{\buildrel {\rm odd} \over {\rm even}}} (\tau) = \sum_m
\gamma_{{\buildrel {\rm odd} \over {\rm even}}} (m) \, {\rm Prob}
(m|t).
\ee
If $\gamma (m)$ is a rapidly decreasing function of $m$, then only small
values of $m$ contribute to $C(\tau)$.  
Thus $C_{DRD}$ varies as $t^{-1/4}$ whenever correlations in the
IS are short ranged. This will be so for most  sectors of the DRD model.
However, we can imagine preparing the system in way such that 
${\tilde \gamma}(k)\/$, the Fourier transform of $\gamma (m)\/$,
varies as $k^{\alpha}\/$ , as 
$|k| \rightarrow 0\/$ with $\alpha$ a given adjustable parameter. 
In such a situation, the decay of $C_{DRD}(\tau)$ with $\tau$ will be 
power law with the exponent depending on $\alpha$.

\section{Hydrodynamic Description of a Many-Sector Decomposable System}

Usually, to describe the behavior of a system for large length or time 
scales, the hydrodynamical description is most appropriate. The number 
of fields that have to be used in such  adescription depends on the 
number of conservation laws in the system. For example, usual 
hydrodynamics involves $5$ densities ( of particle number, momentum and 
energy) corresponding to the fact these are the only locally conserved 
quantities. 
In systems with an infinity of conservation laws, one would expect an 
infinity of such densities, which would make a hydrodynamical 
description impossible. Of course, well-known hydrodynamical equations 
such as the Korteweg de Vries equation are known to have an infinity of 
constants of motion. We would like to understand  what is the 
relationship between existence of infinity of independent constants of 
motion and the hydrodynamical description. Our study of the DRD model 
does not allow us to answer this question fully, but suggests that the 
appearance of {\it arbitrary functions} in the hydrodynamical equations 
of motion is a signal of partial integrability.

In the DRD model,  a hydrodynamic description would have to be in 
terms of coarse-grained density fields $\rho_{odd}(x,t)$ and
$\rho_{even}(x,t)$ where $x$ is a continuous variable 
($0 \leq x\leq L$). These are the local densities of 0's
on odd and even sublattice sites respectively. Equivalently,
we can use the integrated density fields $m_{odd}\/$ and $m_{even}\/$
respectively, where 
\begin{equation}
m_{odd}(x,t) = \int_0^x \rho_{odd}(x^\prime,t) dx^\prime,
\end{equation}
and $m_{even}(x,t)$ is defined similarly. The equations of motion of
$m_{odd}\/$ and $m_{even}\/$ can be transformed into simple linear diffusion
equations characterizing the exclusion process by a non-linear change of
variables from coordinates $x\/$ to coordinates $\xi\/$ (Eqs.30,31).  

The constraint that 
0's do not cross (equivalent to the restriction that the B and C 
particles of the
3-type exclusion process must maintain their
relative ordering),
implies that $m_{odd}(x,t)$ is a known function of
$m_{even}(x,t)$ (and vice versa).  Thus we can write
\begin{equation}
m_{odd}(x,t) = f(m_{even}(x,t)),
\end{equation}
where $f(m)\/$ is a positive, non-decreasing function of its
argument $m\/$, with $f(0) = 0$. Evidently $f\/$ is determined completely
by the irreducible string, as all 0's reside on the IS.
Alternatively, we may determine $f\/$
completely if the initial occupations of each sublattice,
$\rho_{odd}(x,t=0)$ and $\rho_{even}(x,t=0)$ are
known. The function $f(m)$ which is
specified by the initial conditions appears {\it explicitly}
in the equations of motion as a constraint condition.
This equation of constraint is a consequence of the conservation of
the IS.

In a more general context, we can imagine $n\/$ coarse-grained density 
fields $\{\phi\}\/$, whose 
evolution equations are such that after suitable transformations of 
fields, we can define a field $\phi_1$, such that its evolution does 
not depend on the remaining fields, and hence can be written in the form 
\begin{equation}
\frac{\partial \phi_1}{\partial t} = F(\phi_1 ), 
\end{equation}
while the other fields $\{\phi_j\} , j= 2 \ldots n$ may involve couplings 
to $\phi_1$ in addition to couplings to each other. If the evolution 
equation for $\phi_1$ is integrable, we can explicitly
write  $\phi_1(x,t)$ in terms of $\phi_1(x,t=0)$. 
This would imply that the system has an infinity of
conservation laws.  In that
case, the equations of motion of the other fields $\{\phi_j\}\/$,
will involve an {\it arbitrary} function $\phi_1(x,t=0)\/$.
In the DRD model, the equation(38) is a constraint equation involving 
the hydrodynamical fields and an initial condition dependent function.

\section{ Slowly Restoring Ergodicity }

   We shall now try to describe the phenomena of ageing within the 
context of the DRD model. Thus, we would like to describe the system at 
time scales when the slow processes can not be fully ignored. These slow 
processes have rates which decrease sharply with temperature, and are 
negligible at temperatures lower than some `glass temperature'. At larger 
temperatures, these processes would cause transitions between different 
sectors at finite rates, and restore ergodicity.
 
  For the DRD model, the simplest process that causes a transition 
between different sectors ( and thus changes the IS), would be exchanging
neighboring B-  and C- type particles in the IS. This process is however 
not local in the DRD model, as in order to decide if a particular $1$ at 
a site is part of an A- type or B- type particle of the ExP, we have to 
look to the left till a  $0$ is encountered, and this may be many spaces 
away.
    The simplest process which changes the IS, and is local in the DRD model
is
\be
   \ldots0010\ldots \leftarrow\rightarrow \ldots0100\ldots
\ee
In terms of the ExP, this corresponds to the reversible transition
\be
   \ldots CCB\ldots \leftarrow\rightarrow \ldots CBC\ldots
\ee
Let us assume that this process occurs at rate $\Gamma$, which is much less 
than 1. Then for times $t << 1/\Gamma$, we have broken ergodicity and the 
IS is a constant of motion. For times of order $1/\Gamma$, the system 
moves from one sector to another. However, in this case, the ergodicity 
is not completely restored. Of course, in this case $N_B$ and $N_C$ are 
constants of motion. But we are not concerned about $O(L^2)$ decomposition 
of phase space. If $N_C > N_B $, then it is easy to see 
that the $ (N_B + N_C)!/(N_B! N_C!)$ sectors with having same value of
$N_B$ and $N_C$ get connected by the slow dynamics, and ergodicity is 
restored. If however $N_B > N_C$, then the C-type particles can only move
about in a limited region of space, and not be able to go everywhere.
 
  For example, if in the IS , the number of $B$'s between two 
consecutive $C$'s is everywhere at least two, it is easy to see that such 
IS's cannot evolve even under the slow dynamics. The number of such IS's 
is easily seen to grow exponentially with L. Thus this slow process 
merges sectors, but still an exponentially large number of them survive.
The ergodicity is restored only partially.

   Consider an IS with a finite number of $C$'s in a sea of $B$'s. Say 
the configuration is $\ldots BBBCCCBBBB\ldots$. Then it is easy to see 
that the leftmost $C$ cant move at all, and the rightmost $C$ by at most 
two characters to the right. In general, a configuration with n consecutive 
$C$'s can 
spread out so that there are no adjacent $C$'s, leaving the leftmost $C$ 
fixed, and no further.

   It is straight forward to count the number of sectors in the presence 
of the slow process. For each of these big sectors, we can get a standard
configuration by by pushing all $C$'s as far to the right as possible. 
Then its structure is a random string formed with the substrings $CB$ and
$B$ in arbitrary order. It is easily seen that this number is 
$N_B!/[N_C!(N_B-N_C)!]$, which grows exponentially with  the size of 
system.  
    To study relaxation of auto-correlations in the presence of this slow 
relaxation is quite straight forward. For times $t$ much less than 
$1/\Gamma$, the autocorrelation typically decays as $t^{-1/4}$. However 
for larger times in case ergodicity is restored fully, it will crossover 
to a diffusive $t^{-1/2}$ decay. In case $N_B > N_C$, the decay would 
continue to 
be $t^{-1/4}$ for large times, but presumably with a modified coefficient.
Further study of this model is in progress.

\section{ Generalizations and Relation to other models}

In summary, the DRD model is  a very simple model for studying  dynamics of 
MSD systems i.e. those systems showing strong breaking of  ergodicity, 
in which the phase space can be decomposed into an exponentially large
number of mutually disconnected sectors. For this model, we can determine 
the sizes and
numbers of these exactly.  The sectors can be labelled 
 by different values of a conserved quantity, the
Irreducible String.  The exact equivalence of the model to a model of
diffusing hard core random walkers with conserved spin allowed us to
determine the sector-dependent behaviour of time-dependent correlation
function in different sectors.  In any given sector, we showed that
the stochastic rate matrix was the same as that of the quantum Hamiltonian 
of a spin$-1/2$
Heisenberg chain (whose length depends on the sector), and thus demonstratated
that it was exactly diagonalizable.

The construction of the irreducible string in this model is
very similar to other one-dimensional stochastic models with the
many-sector-decomposability property studied recently, i.e. the $k$-mer
deposition-evaporation model \cite{bar1,bar2,china} and the $q$-color dimer
deposition-evaporation model \cite{harimenon,thesis}.
In all of these models, the long-time behaviour of the
time-dependent correlation functions can be determined by noting  
that it is expected to behave
qualitatively in the same way as that of the spin-spin autocorrelation 
function in the HCRWCS model with the same spin sequence as the IS in
the corresponding sector. 

However, the DRD model differs from earlier studied models in
significant ways.  In the trimer deposition evaporation model (TDE),
the correspondence between the configurations on the line and the position
of hard core walkers is many to one, unlike the present model where it
is one-to-one.  As a consequence, in the steady state of the TDE model,
all configurations of the random walkers are not equally likely, and
one has to introduce an effective interaction potential between the
walkers which is found to be of the form
\be
V(\{X_i\}) = \sum_i f(X_{i+1} - X_i)
\ee
where $\{X_i\}$ are the positions of the walkers, and $f(x)$ increases as
${3 \over 2} \ln X$ for large $X$.

In the TDE model, the transition probabilities for the random
walkers are also not completely independent of the spin-sequence of the
walkers.  In the $q$-color dimer deposition evaporation (qDDE) model,
the color symmetry of the model implies that the dynamics of random
walkers is completely independent of the spin sequence of the walkers,
but the potential of interaction $V$ is still present, which makes the
problem difficult to study exactly.  The DRD model is thus simpler
than both the TDE and the qDDE models, and has the additional virtue
of being exactly solvable in the sense that the stochastic matrix can
be diagonalized completely.

\begin{figure}
\centerline{
        \epsfxsize=8.0cm
        \epsfysize=10.0cm
        \epsfbox{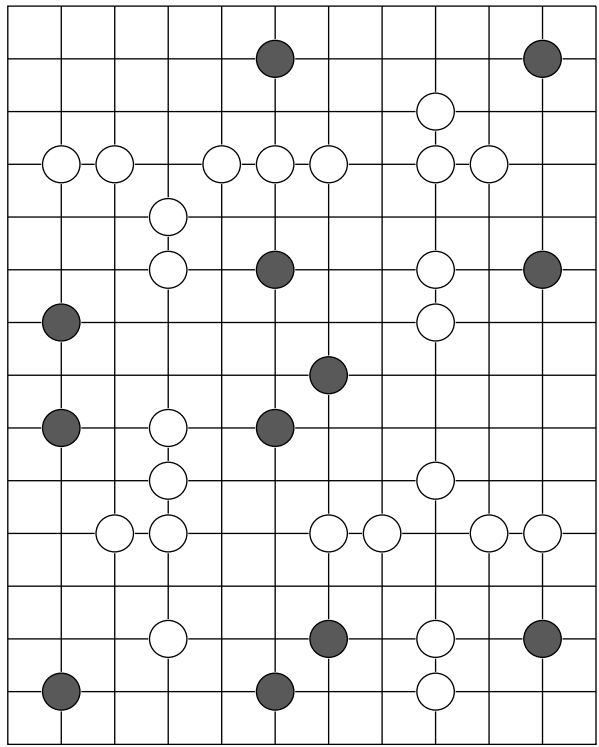}
}
\caption{ An example of a partially stuck configuration of the DRD model 
in two dimensions. The particles denoted by black circles are completely 
immobile, and cannot move under the DRD rules. The mobile particles ( 
denoted by white circles)can do so only along some horizontal or vertical 
lines.} 
\label{fig:3} 
\end{figure}


There are some straightforward but interesting generalizations of
the model.  Consider a general exclusion process with $k$ types of
particles.  In this general model, if we assume that some types of
particles cannot exchange positions (setting their exchange rate to
zero), their relative order will be conserved and this
can be coded in terms of the conservation of an IS.
As a simple example, consider a model with 4 species of particles
labelled A,B, C and D respectively with the allowed exchanges with
equal rates
\be
\begin{array}{l}
AC \rightleftharpoons CA, \;\;\; AD \rightleftharpoons DA \\
BC \rightleftharpoons CB  \;\;\; BD \rightleftharpoons DB.
\end{array}
\ee
This model again has the MSD property.  It
is easy to see that there are now {\it two} irreducible strings
which are conserved by the dynamics.  This is because the dynamics
conserves the relative order of A and B type particles, and also
of C and D type particles.  In a string specifying the
configuration formed of letters A,B,C and D, deleting all occurrences
of A and B gives rise to an IS specifying the relative
order between C and D type particles, which is a constant of motion.
Similarly, deleting all occurrences of C and D characters, we get
another independent IS which is also a constant of
motion.  In a specific sector, where both irreducible strings are
known, the dynamics treats A and B particles as indistinguishable,
as also  C and D.  Thus the dynamics is the same as that of the simple
exclusion process (with only two species of particles), and is
equivalent to the exactly solved Heisenberg model.

 The DRD model is  interesting in higher dimensions also .  For
example, it is easy to see that on a square lattice in two dimensions,
the number of totally jammed configurations increases
exponentially with the number of sites in the system.  All
configurations with no two adjacent 1's are totally jammed.  These are
just the configurations of the hard-square lattice gas model\cite{baxter},
whose number is known to increase exponentially with the area of
the system.  One can also construct configurations in which almost all
sites are jammed, except for a finite number of diffusing dimers,
which move along a finite set of horizontal or vertical lines (see fig. 
3). Thus the number of sectors in which only some of the particles are 
moving also increases as exponential of the volume of the system. However, 
there is no equivalence to the 2-d Heisenberg model, and no obvious 
analog of the IS to label the disjoint sectors, and solving for the 
dynamics seems quite difficult. 

An interesting generalization of the model would be to the case when the 
hamiltonian is non-trivial,  say an Ising model with nearest or nearest 
and next-nearest neighbor couplings. Then the allowed processes are 
assumed to be same as in the simple DRD model discussed above, but the 
rates are assumed to satisfy detailed balance. In this case, in the 
steady state, realtive weights of different configurations are given by 
the Boltzmann weights, and one can determine the pico-canonical partition 
functions as functions of parameters like temperature in any given sector. 
The relative 
weights of different sectors in the sector-averaging define another 
temperature parameter in the system. This may be identified as the 
temperature  at which freezing occurs( quenching 
temperature). Details will be discussed  in a future 
publication.

Another generalization of the model would be to make the diffusion
asymmetric.  The corresponding 3-species exclusion process then
becomes asymmetric, and belongs to the KPZ universality class\cite{KPZ}.  The
invariance of dynamics under exchanging B and C type particles DD again 
reduces the problem  to a simple
asymmetric exclusion process of 2 species, known to be exactly soluble
by Bethe ansatz techniques \cite{gwaspohn,halpin,kim},
and has a non-classical dynamical exponent 3/2.  The correlation
function of the asymmetric DRD model would map to somewhat complicated 
multispin correlation functions of the simple asymmetric exclusion 
process.  How these would vary from sector to sector has not been 
studied so far.

One intriguing feature of this model ( and other models with conserved 
IS) is the fact that we are able to get an infinity of classical 
constants of motion by constructing a one parameter family of matrices 
which commute with each other and with the Hamiltonian. A similar 
construction is used in integrable models, involving the R-matrix which 
is assumed to satisfy the Yang -Baxter equation. The similarity of 
construction suggests that one can devise a more general R-matrix from 
which both the classical and quantum mechanical conserved quantities are 
deducible. Such a structure remains to be found. 

     Acknowledgements: It is a pleasure to thank M. Barma, H. M. 
Koduvely and G. I. Menon for many useful discussions, and A. Dhar for 
help in preparation of the manuscript.


\begin{thebibliography}{99}
\bibitem{footnote1} The words `microcanonical' and `canonical' have been 
used here in a somewhat more general sense than in conventional 
statistical mechanics textbooks. We use the former to mean the case when 
the partition function is an integral over phase space, with some product 
of delta functions as integrand. In the latter case, the integrand is of 
the form $exp(-a_1O_1 -a_2O_2 \ldots)$, where $O_1, O_2$ are some 
observables, and $a_1, a_2 $ are some parameters.
\end{thebibliography}
\end{document}